\documentclass[a4paper,11pt]{article}
\usepackage[utf8x]{inputenc}
\usepackage{amsmath}
\usepackage[T1]{fontenc}
\usepackage{authblk}
\usepackage{ulem}

\usepackage{dcolumn}
\usepackage{bm}
\usepackage{slashed}
\usepackage[toc,page]{appendix}
\usepackage{float}
\usepackage{graphicx,xspace,units}
\usepackage{float}
\usepackage{amsmath}
\usepackage{amssymb}
\usepackage{color}

\usepackage{graphicx}

\title{ 
\bf \boldmath 
Constraints on WIMP Annihilation for Contracted Dark Matter in the Inner Galaxy with the \textit{Fermi}-LAT
\\[8mm]}

\author[1,2,3]{Germ\'{a}n~A.~G\'{o}mez-Vargas}
\author[4]{Miguel A. S\'anchez-Conde}
\author[1,2,5]{Ji-Haeng Huh\thanks{MultiDark Fellow}}
\author[1,2]{Miguel Peir\'o\thanks{MultiDark Scholar}}
\author[6,2,7]{Francisco Prada}
\author[3]{Aldo Morselli}
\author[8]{Anatoly Klypin}
\author[1,2]{David G. Cerde\~{n}o}
\author[9]{Yann Mambrini}
\author[1,2]{Carlos~Mu\~{n}oz}
\affil[1]{{\small Departamento de F\'{\i}sica Te\'{o}rica, Universidad Aut\'{o}noma de Madrid, Cantoblanco, E-28049 Madrid, Spain}}
\affil[2]{{\small Instituto de F\'{\i}sica Te\'{o}rica UAM--CSIC, Campus de Cantoblanco UAM, 28049 Madrid, Spain}}
\affil[3]{{\small Istituto Nazionale di Fisica Nucleare, Sez. Roma Tor Vergata, Roma, Italy}}
\affil[4]{{\small SLAC National Accelerator Laboratory \& Kavli Intitute for Particle Astrophysics and Cosmology, 2575 Sand Hill Road, Menlo Park CA, 94025, USA}}
\affil[5]{{\small Department of Physics and Astronomy, UCLA, 475 Portola Plaza, Los Angeles, CA 90095, USA}}
\affil[6]{{\small Campus of International Excellence UAM/CSIC, Cantoblanco, 28049 Madrid, Spain}}
\affil[7]{{\small Instituto de Astrof\'{\i}sica de Andaluc\'{\i}a, Glorieta de la Astronom\'{\i}a, 18008 Granada, Spain}}
\affil[8]{{\small Astronomy Department, New Mexico State University, Las Cruces NM, USA}}
\affil[9]{{\small Laboratoire de Physique Th\'eorique Universit\'e Paris-Sud, F-91405 Orsay, France}}

\begin{document}






\maketitle

\begin{abstract}
We derive constraints on parameters of generic dark matter candidates by comparing theoretical predictions with the gamma-ray emission observed by the \textit{Fermi}-LAT from the region around the Galactic Center. 
Our analysis is conservative since it simply requires that the expected dark matter signal does not exceed the observed emission.
The constraints obtained in the likely case that the collapse of baryons to the Galactic Center is accompanied by the contraction of the dark matter are strong.
In particular, we find that for $b\bar b$ and $\tau^+\tau^-$ or $W^+W^-$ dark matter annihilation channels, the upper limits on the annihilation cross section imply that the thermal cross section is excluded for 
a Weakly Interacting Massive Particle (WIMP) mass smaller than about 700 and 500 GeV, respectively.  
For the $\mu^+ \mu^-$ channel, where the effect of the inverse Compton scattering is important, depending on models of the Galactic magnetic field the exclusion of the thermal cross-section is for a WIMP mass smaller than about 150 to 400 GeV.
The upper limits on the annihilation cross section of dark matter particles obtained are two orders of magnitude stronger than without contraction.
 In the latter case our results are compatible with the upper limits from the Galactic halo analysis reported by the \textit{Fermi}-LAT collaboration for the case in which the same conservative approach without modeling of the astrophysical background is employed.

\end{abstract}

 \section{Introduction}

 
Astrophysical searches for dark matter (DM) are a fundamental part of the experimental efforts to explore the dark sector.
The strategy is to search for DM annihilation products in preferred
regions of the sky, i.e., those with the highest expected DM
concentrations and still close enough to yield high DM-induced fluxes
at the Earth. For that reason, the Galactic Center (GC), nearby dwarf
spheroidal galaxy (dSphs) satellites of the Milky Way, as well as
local galaxy clusters are thought to be among the most promising
objects for DM searches. In particular, dSphs represent very
attractive targets because they are highly DM-dominated systems and
are expected to be free from any other astrophysical gamma-ray emitters
that might contaminate any potential DM signal. Although the expected
signal cannot be as large as that from the GC, dSphs 
may produce a larger signal-to-noise ($S/N$) ratio. This fact
allows us to place very competitive upper limits on the gamma-ray
signal from DM annihilation
\cite{Ackermann:2011wa, Abdo:2010, GeringerSameth:2011iw}, using data
collected by the Large Area Telescope (LAT) onboard the Fermi
gamma-ray observatory \cite{Atwood:2009ez}. These are often referred to
as the most stringent limits on DM annihilation cross-section obtained so far.

Despite these interesting limits derived from dSphs, the GC is still expected to be the brightest
source of DM annihilations in the 
gamma-ray sky by several orders of magnitude. Although several
astrophysical processes at work in the crowded GC region make it extremely difficult to
disentangle the DM signal from conventional emissions, the DM-induced 
gamma-ray emission is expected
to be so large there that the search is still worthwhile. Furthermore, the DM density in the GC may be larger than what is typically
obtained in $N$-body cosmological simulations. Ordinary matter
(baryons) dominates the central region of
our Galaxy  \cite{Prada:2004pi}. Thus, baryons may significantly affect the DM distribution. As
baryons collapse and move to the center they increase the
gravitational potential, which in turn forces the DM to
contract and increase its density.  This is a known and qualitatively
well understood physical process
\cite{Zeldovich,blumenthal,gnedin04}. It is also observed in many
cosmological simulations that include hydrodynamics and star
formation \cite{gustafsson,colin06,tissera10,gnedin11,Zemp:2011nk,SommerLarsen:2009me}. If this is the
only effect of baryons, then the expected annihilation signal will
substantially increase \cite{Prada:2004pi,mambrini}.



In this work, we analyze in detail the
constraints that can be obtained for generic DM candidates from
\textit{Fermi}-LAT inner Galaxy gamma-ray measurements assuming some specific (and well motivated)
DM distributions.
The approach is conservative, requiring
simply that the expected DM signal does not exceed the gamma-ray emission observed by the \textit{Fermi}-LAT in an optimized region around the
GC. The region is chosen in such a way that 
the $S/N$ ratio is maximized.
This kind of analysis, without modeling of the astrophysical background, was also carried out by
the \textit{Fermi}-LAT collaboration to constrain DM models from Galactic halo observations \cite{fermidiffuse}.
%
%

The paper is organized as follows. In Section \ref{profiles} we
discuss DM density profiles such as Navarro, Frenk and White (NFW) \cite{nfw96,nfw97}, Einasto
\cite{einasto,navarro04} and Burkert \cite{burkert}, paying special attention to the effect
of baryonic compression.  In Section \ref{flux} we model the gamma-ray flux from DM annihilation, taking into account the
contributions from prompt photons and photons induced via inverse
Compton scattering (ICS). The latter is specially
relevant for the $\mu^+ \mu^-$ channel.  We will see that compressed
profiles significantly increase the gamma-ray flux in the inner
regions of the Galaxy.
In Section \ref{data} we analyze the gamma-ray flux from
\textit{Fermi}-LAT measurements. For that we use an optimized region
around the GC, which will depend on the particular DM density profile
considered.
Taking into account these
results, in Section \ref{limits} we are able to obtain significant
limits on the annihilation cross-section for a generic DM candidate
annihilating to $b\bar b$, $\tau^+\tau^-$, $\mu^+ \mu^-$ or $W^+W^-$,
in the case that the collapse of baryons to the GC is accompanied by the
contraction of the DM. 
In general,
the upper limits on the annihilation cross section of DM particles are two orders of magnitude stronger than without contraction, where the
thermal cross section is not excluded.
Finally, the conclusions are left for Section \ref{conclusions}.



 \section{\bf Dark matter density profiles}
\label{profiles}


Cosmological $N$-body simulations provide important results regarding
the expected DM density in the central region of our
Galaxy. Simulations suggest the existence of a universal DM
density profile, valid for all masses and cosmological epochs. It is
convenient to use the following parametrization for the DM halo
density \cite{kravtsov}, which covers different approximations for DM density:
\begin{equation}
\rho(r) = \frac{\rho_s}{\left(\frac{r}{r_s}\right)^\gamma \left[1 +\left(\frac{r}{r_s}\right)^\alpha\right]^
{\frac{\beta - \gamma}{\alpha}}}\ , 
\label{eq:kravtsov}
\end{equation}
where $\rho_s$ and $r_s$ represent a characteristic density and a scale radius, respectively. 
The NFW density profile \cite{nfw96,nfw97}, with
($\alpha$,$\beta$,$\gamma$) = (1,3,1), is by far the most widely used
in the literature. 
Another approximation is  the so-called
Einasto profile \cite{einasto,navarro04}
\begin{equation}
\rho_{\text{Ein}}(r)=\rho_s \exp \left\{-\frac{2}{\alpha}\left[\left(\frac{r}{r_s}\right)^{\alpha} - 1\right] \right\}\ ,
\label{eq:einasto}
\end{equation}
which provides a better fit than NFW to numerical results
\cite{navarro04,Merritt}. Finally, we will also consider DM density profiles that possess a core at the center, such as the purely phenomenologically motivated Burkert profile \cite{burkert}:

\begin{equation}
\rho_{\text{Burkert}}(r)=\frac{\rho_s~r_s^3}{(r+r_s)~(r^2+r_s^2)}\ .
\label{eq:burkert}
\end{equation}

Early results on the central slopes of the DM profiles showed
some significant disagreement between the estimates, with values ranging from $\gamma =1.5$
\cite{Moore:1999} to $\gamma =1$ \cite{nfw96,nfw97}.  As the accuracy
of the simulations improved, the disagreement became smaller. For the
Via Lactea II (VLII) simulation the slope in Ref.~\cite{VLII} was
estimated to be $\gamma=1.24$. A re-analysis of the VLII simulation and
new simulations performed by the same group give the slope
$\gamma=0.8-1.0$ \cite{Stadel}, which is consistent with the Aquarius
simulation \cite{springel08}. Another improvement comes from the fact
that the simulations now resolve the cusp down to a radius of $\sim
100$~pc, which means that less extrapolation is required for the density of the central region.

Yet, there is an additional ingredient
that is expected to play a prominent role in the centers of DM halos:
baryons. Although only a very small fraction of the total matter
content in the Universe is due to baryons, they represent the
dominant component at the very centers of galaxies like the Milky
Way. Actually, the fact that current N-body simulations do not resolve
the innermost regions of the halos 
is a minor consideration relative to the uncertainties due to
the interplay between baryons and DM. 

The baryons lose energy through
radiative processes and fall into the central regions of a forming
galaxy. As a consequence of this redistribution of mass, the resulting
gravitational potential is deeper, and the DM must move closer to the
center, increasing its density.
This {\it compression} of DM halos due to baryonic infall was first
studied in Ref.~\cite{Zeldovich} for a spherically symmetric DM halo
using simple simulations and adiabatic invariants. A convenient
analytical approximation was provided in Ref.~\cite{blumenthal}. The
model was later modified \cite{gnedin04} to account for the
eccentricity of orbits of DM particles.  The effect seems to
be confirmed by recent hydrodynamic simulations (see
e.g.~Refs.~\cite{gustafsson,colin06,tissera10,gnedin11,Zemp:2011nk,SommerLarsen:2009me}.). In
Ref.~\cite{gnedin11}, for instance, the authors ran a set of
high-resolution hydrodynamic simulations that self-consistently
included complex baryonic physics such as gas dissipation, star
formation and supernova feedback. They all showed a clear steepening
of the inner DM density profiles with respect to DM-only
simulations. Indeed, it is argued by the authors that such effect
should be always included in order to correctly model the mass
distribution in galaxies and galaxy clusters.
 
As pointed out in Ref.~\cite{Prada:2004pi}, the effect of the baryonic
adiabatic compression might be crucial for indirect DM searches, as it
increases by several orders of magnitude the gamma-ray flux from DM
annihilation in the inner regions, and therefore the DM
detectability. In Ref.~\cite{mambrini}, this effect was used to study
the detection of supersymmetric DM by the \textit{Fermi}-LAT, with the
conclusion that fluxes from the GC would be largely reachable in
significant regions of the supersymmetric parameter space. The effect
of compression on galaxy clusters was recently studied in
Ref.~\cite{andoclusters}.

There is however another possible effect related to baryons
that tends to decrease
the DM density and flatten the DM cusp \cite{Mashchenko:2006, Mashchenko:2008,Pontzen}.
The mechanism relies
on numerous episodes of baryon infall followed by a strong burst of
star formation, which expels the baryons. At the beginning of each
episode the baryons dominate the gravitational potential. The DM 
contracts to respond to the changed potential. A sudden onset of
star formation drives the baryons out. The DM also moves
out because of the shallower potential. Each episode produces a relatively
small effect on the DM, but a large number of them results in a significant
decline of the DM density.
Indeed, cosmological simulations
that implement this process show a strong decline of the DM
density \cite{Governato, Maccio}. 
Whether the process happens in reality is
still unclear. Simulations with the cycles of
infall-burst-expansion process require that the gas during the burst
stage does not lose energy through radiation, which is not
realistic. Still, the strong energy release needed by the mechanism
may be provided by other processes and the flattening of the DM
cusp may occur. 
If this happened to
our Galaxy, then the DM density within the central $\sim 500$~pc may
become constant \cite{Maccio}. This would reduce the annihilation
signal by orders of magnitude. We note that this mechanism would
wipe out the DM cusp also in centers of dwarf galaxies
\cite{Governato}. 
Yet, a recent work that also includes 
stellar feedback offers a much more complicated 
picture in which galaxies may retain or not their DM cusps depending on the ratio between their stellar-to-halo masses \cite{dicintio13}.

\begin{table}
\centering
\begin{tabular}{|c|c|c|c|c|c|c}
  \hline
   Profile \rule{0pt}{3ex} &  $\alpha$  &   $\beta$  & $\gamma$ &   $\rho_s$ [GeV cm$^{-3}$] &     $r_s$ [kpc]          \\
   \hline \rule{0pt}{3ex}
   Burkert  \rule{0pt}{3ex}&    $--$   &     $--$   & $--$     &  $37.76$            &     $2$           \\   
   Einasto &    $0.22$   &     $--$   & $--$     &  $0.08$            &     $19.7$           \\    
   NFW  &    $1$      &    $3$     & $1$      &  $0.14$            &     $23.8$           \\
   NFW$_c$  &    $0.76$   &    $3.3$   & $1.37$   &  $0.23$            &     $18.5$           \\

 \hline
\end{tabular}

\caption{\label{tab:table1} DM density profiles used in this work, 
following the notation of Eqs.~(\ref{eq:kravtsov}-\ref{eq:burkert}).
}
\end{table}


As discussed in the Introduction, in this work we pay special attention to those scenarios where
the DM cusp is not flattened.
However, in order to quantify the uncertainty in the DM density profile, we will use three well-motivated models: Einasto, NFW and a compressed NFW (NFW$_c$), whose parameters have been constrained from observational data of the Milky Way, as well as a cored Burkert profile, also compatible with current constraints.
We have followed Ref.~\cite{Prada:2004pi} to choose the parameters of both the NFW and the NFW$_c$. We have fitted the resulting data of that work with the power-law parametrization of Eq.~(\ref{eq:kravtsov}).
The results for both profiles are listed in Table~\ref{tab:table1}. The effect of baryonic adiabatic compression is clearly noticed
 at small $r$ as a steep power law $\rho\propto 1/r^{\gamma}$ with $\gamma=1.37$ for NFW$_c$, which is in contrast to the standard
  NFW value, $\gamma=1$. We note that a value of $\gamma=1.37$ is indeed perfectly consistent with what has been found in recent
   hydrodynamic simulations \cite{gnedin11} and it is also compatible with current observational constraints (mainly derived from
  microlensing and dynamics) on the slope of the DM density profile in the central regions of the Milky Way \cite{iocco11}. These studies actually allow for even steeper adiabatically contracted profiles. Concerning the Einasto profile we will use the parameters provided in Ref.~\cite{Catena:2009mf}. Finally, for the Burkert profile, we decided to choose a core radius of 2 kpc. This core size is in line with previous works \cite{fermidiffuse,kelso} and with that suggested by recent hydrodynamic simulations of Milky Way size halos \cite{eris}. For the normalization of the profile we chose the value of the local density suggested in Ref.~\cite{Catena:2009mf} for Milky Way Burkert-like profiles. The resulting profile is also compatible with current observational constraints \cite{iocco11}. Note, however, that a recent work favors a substantially larger core radius and a slightly higher normalization for Burkert-like profiles \cite{salucci13}. All the profile parameters are summarized in Table~\ref{tab:table1} and the four profiles are shown in the left panel of Figure~\ref{fig:J}.
 
Let us finally point out that 
there are other possible effects driven by baryons that might steepen the DM density profiles in the centers of DM halos, such as
central black holes (see e.g. Refs.~\cite{gondolo,gnedinprimack,bertone02}), that we will not consider here. 

 
  \section{Gamma-ray flux from dark matter annihilation}
  
\label{flux}


The gamma-ray flux from DM annihilation in the Galactic halo has two main contributions \cite{bernal}: prompt photons 
and photons induced via ICS.
The former are produced indirectly through hadronization, fragmentation and decays of the DM annihilation products or by
internal bremsstrahlung, or directly through one-loop processes (but these are typically suppressed in most DM models). 
The second contribution originates from electrons and positrons produced in DM annihilations, via ICS off the ambient photon background.
The other two possible contributions to the gamma-ray
flux from DM annihilation can be neglected in our analysis: radiation
from bremsstrahlung is expected to be sub-dominant with respect to ICS
in the energy range considered ($1$ - $100$ GeV) and a few
degrees off the Galactic plane (see Fig. 14 in Ref.~\cite{FermiLAT:2012aa}), and
synchrotron radiation is only relevant at radio frequencies, below the
\textit{Fermi}-LAT threshold.
Thus the gamma-ray differential flux from DM annihilation 
from a given observational region $\Delta\Omega$ in the Galactic halo can be written as follows:
\begin{eqnarray}
\frac{d\Phi_\gamma}{dE_\gamma} (E_\gamma,\Delta\Omega)&=&
\left(\frac{d\Phi_\gamma}{dE_\gamma}\right)_{prompt}+
\left(\frac{d\Phi_\gamma}{dE_\gamma}\right)_{ICS}\ .
\label{eq1}
\end{eqnarray}
We discuss in detail both components in the next subsections.

\subsection{Prompt gamma rays}

A continuous spectrum of gamma rays is produced mainly by the decays of $\pi^0$'s generated in the cascading of annihilation products and also by internal bremsstrahlung. 
While the former process is completely 
determined for each given final state of annihilation (we will study $b\bar b$, $\tau^+\tau^-$, $\mu^+ \mu^-$ and $W^+W^-$ channels), the latter depends in general on the details of the DM model such as the DM particle spin and 
the properties of the mediating particle. Nevertheless, it is known that  internal bremsstrahlung always includes much model-independent final state radiation (FSR), which is 
emitted directly from charged particles in the external legs \cite{Birkedal:2005ep,Bringmann:2007nk}. 
In our analysis of generic DM models, 
we only consider these FSR
components of the internal bremsstrahlung.
It is a safe choice for the conservative approach that we follow, 
since the inclusion of model-dependent emission from virtual charged mediators would make constraints only 
stronger \cite{Bringmann:2007nk,Bringmann:2012vr}.

As we will consider throughout this work the case of self-conjugated DM particles, the prompt contribution can be written as
\begin{eqnarray}
\left(\frac{d\Phi_\gamma}{dE_\gamma}\right)_{prompt}=\sum_i 
\frac{dN^i_{\gamma}}{dE_\gamma}\ \frac{\langle\sigma_i v\rangle}{8\pi m_{DM}^2}\ \bar J (\Delta\Omega)\Delta\Omega\ .
  \label{gammaflux}
\end{eqnarray}
This equation has to be multiplied by an additional factor of $1/2$ if the DM particle studied is not its own anti-particle. The discrete sum is over all DM annihilation channels. $dN^i_{\gamma}/{dE_\gamma}$ is the differential gamma-ray yield\footnote{ For the spectra of gamma rays we use pre-evaluated tables in \cite{Cirelli:2010xx}, which are generated using {\tt PYTHIA} \cite{pythia} and thus containing FSR properly.}, 
$\langle\sigma_i v\rangle$ is the annihilation cross-section averaged over the velocity distribution of the DM particles, 
$m_{DM}$ is the mass of the DM particle, and the quantity $\bar J (\Delta\Omega)$ (commonly known as the {\it J-factor}) is defined as
\begin{eqnarray}
\bar J (\Delta\Omega)\equiv \frac{1}{\Delta\Omega}\int~d\Omega~\int_{l.o.s.} \rho^2(r(l,\Psi))~dl\ .
  \label{gammaflux2}
\end{eqnarray}
The J-factor accounts for both the DM distribution and the geometry of the system\footnote{Although in principle the point-spread function (PSF) should be included in this formula (see e.g., Refs.\cite{Prada:2004pi,masc07,bringmann09}), it turns out to be not relevant in our study mainly for two reasons: i) we deal with fluxes integrated in large regions of the sky, much larger than the PSF, and ii) we avoid the very center of the Galaxy, where the PSF would artificially smear out the cusps expected from some of the DM density profiles. See section \ref{data} for more details.}.
The integral 
of the square of the DM density $\rho^2$ in the direction of observation $\Psi$ is along the line of sight ({\it l.o.s}), and $r$ and $l$ represent the 
galactocentric distance and the distance to the Earth, respectively.
Indeed, in Eq.~(\ref{gammaflux}), all the dependence on astrophysical parameters is encoded in the J-factor itself,
whereas the rest of the terms encode the particle physics input\footnote{ Strictly speaking, both terms are not completely independent of each other, as the minimum predicted mass for DM halos is set by the properties of the DM particle and is expected to play an important role also in the J-factor when substructures are taken into account. In our work, we do not consider the effect of substructures on the annihilation flux, as large substructure boosts are only expected for the outskirts of DM halos \cite{pinzke11,gao12}, and thus they should have a very small impact on inner Galaxy studies.}.
The most crucial aspect in the calculation of $\bar J (\Delta\Omega)\Delta\Omega$ 
is related to the modeling of the DM distribution in the GC.

\begin{figure}[!t]
\centering
\includegraphics[width=0.49\textwidth]{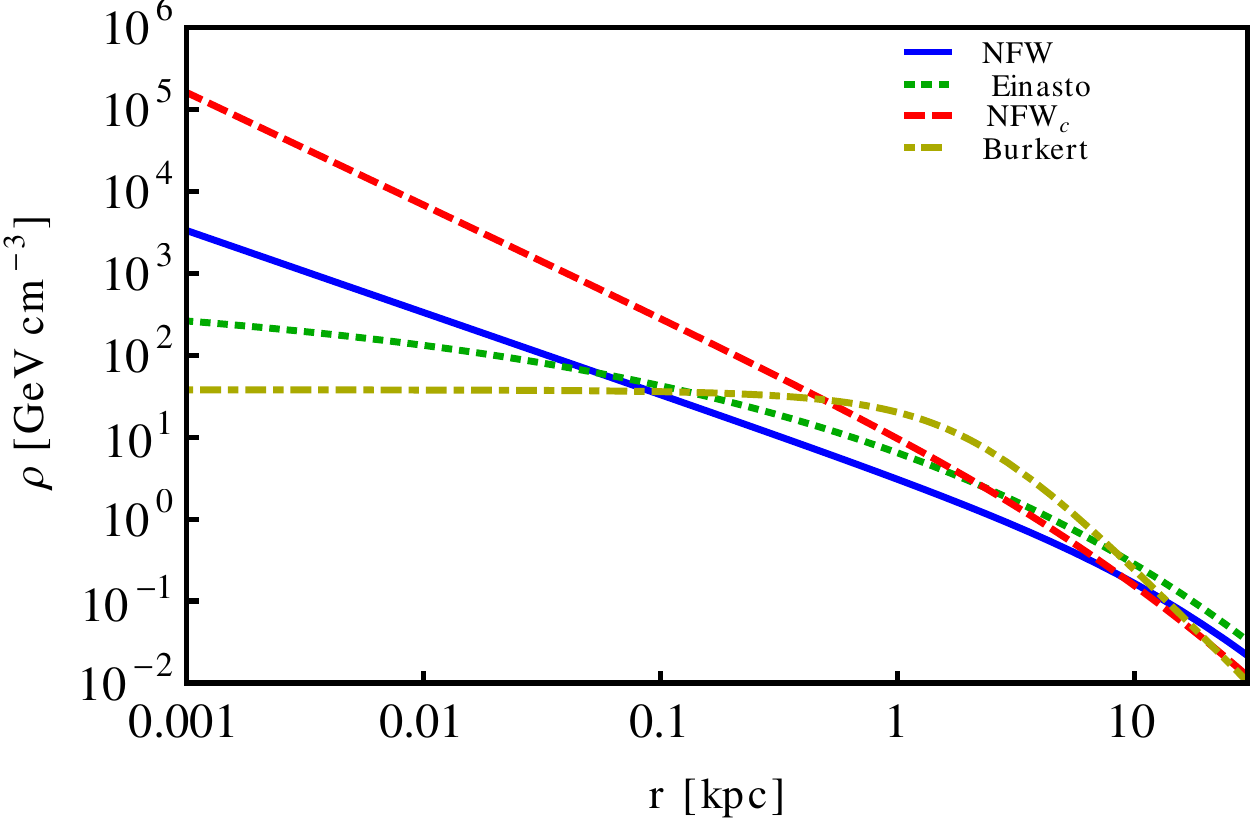}
\includegraphics[width=0.49\textwidth]{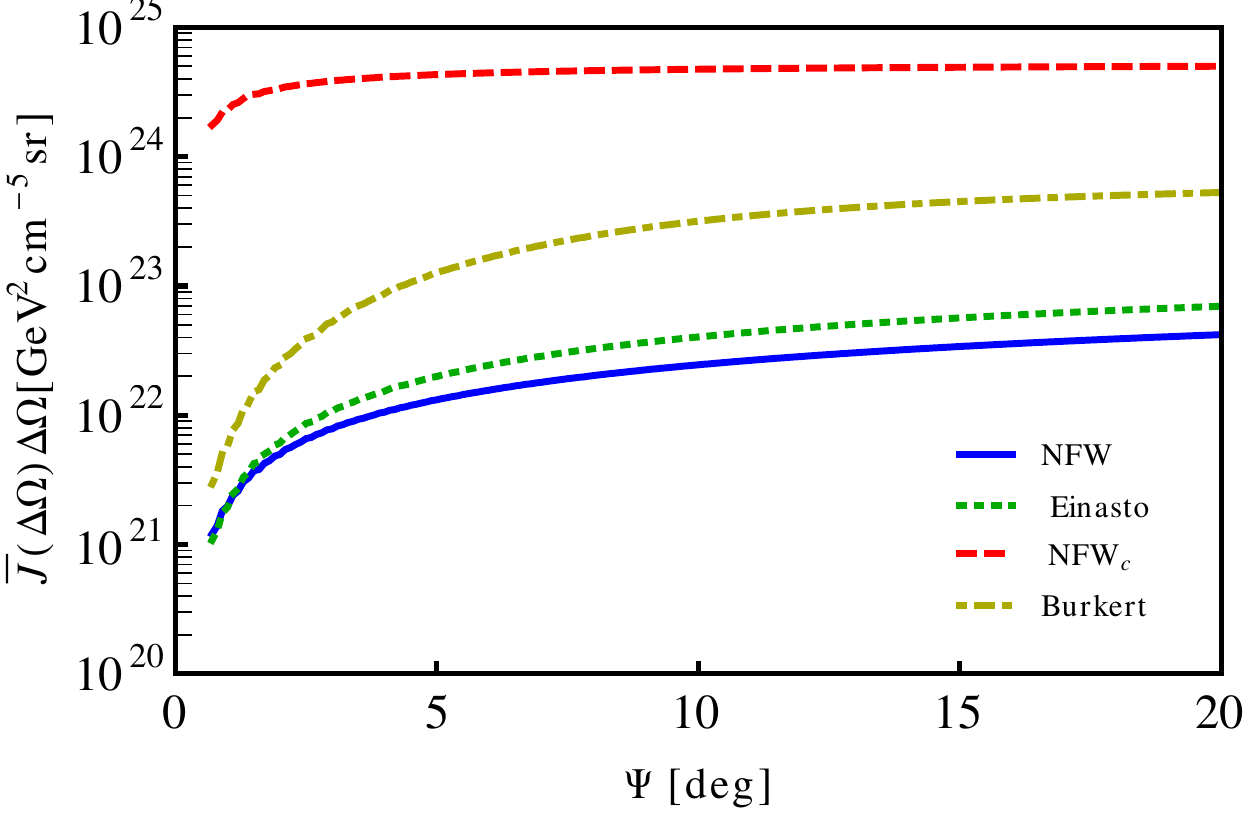}
	 \caption{\label{fig:J} 
 Left panel: DM density profiles used in this work, with the parameters given in Table~\ref{tab:table1}. Right panel: The $\bar J (\Delta\Omega)\Delta\Omega$ quantity integrated on a ring with inner radius of $0.5^{\circ}$ ($\sim 0.07$ kpc) and external radius of $\Psi$ ($R_\odot \tan \Psi$)
for the DM density profiles given in Table~\ref{tab:table1}. Blue (solid), 
red (long-dashed), green (short-dashed) and yellow (dot-dashed) lines correspond to NFW, NFW$_c$, Einasto and Burkert profiles, respectively. The four DM density profiles are compatible with current observational data.} 
\end{figure}

In the right panel of Figure~\ref{fig:J}, the $\bar J (\Delta\Omega)\Delta\Omega$ quantity corresponding to each of the four profiles discussed in Section
\ref{profiles} is shown as a function of the angle $\Psi$ from the GC. 
The associated observational regions $\Delta\Omega$ to each $\Psi$ are taken around the GC.
The angular integration is over a ring with inner radius of $0.5^{\circ}$  and external radius of $\Psi$. 
We have assumed a $r=0.1$ pc constant density core for both NFW and NFW$_c$, although as discussed e.g. in Refs.~\cite{fornengo,mambrini} the results are almost insensitive to any core size below $\sim$1 pc (or even larger given the \textit{Fermi}-LAT PSF).
Remarkably, the adiabatic compression increases the DM annihilation flux by several orders of magnitude in the inner regions, i.e., the regions of interest in the present study. This effect will turn out to be especially relevant when deriving limits on the DM annihilation cross section.
We also note that for the Burkert profile the value of 
$\bar J (\Delta\Omega)\Delta\Omega$ is larger than for the NFW and Einasto profiles.
This is so because of the relative high normalization used for this profile compared to the others and, especially, due to the annular region around the GC where we are focusing our studies, which excludes the GC itself (where such cored profiles would certainly give much less annihilation flux compared to cuspy profiles, see left panel of Figure~\ref{fig:J}). 
We note, however, that the use of another Burkert-like profile with a larger DM core than the one used here, as e.g., the one recently proposed in Ref.~\cite{salucci13}, may lead to substantially lower $\bar J (\Delta\Omega)\Delta\Omega$ values, and thus to weaker DM constraints.
In particular, we checked for the profile in Ref.~\cite{salucci13} that the values of $\bar J (\Delta\Omega)\Delta\Omega$ in the region shown in Figure~\ref{fig:J} are always smaller than about 
$10^{22}$ GeV$^2$ cm$^{-5}$ sr.
Notice finally that the NFW$_c$ profile reaches a constant value of $\bar J (\Delta\Omega)\Delta\Omega$ for a value of $\Psi$ smaller than the other profiles. This is relevant for our discussion below for optimization of the region of interest for DM searches, since we see that for NFW$_c$ a larger region of analysis will not increase the DM flux significantly as for NFW, Einasto and Burkert profiles.

         \subsection{Gamma rays from Inverse Compton Scattering}
         
         \label{ICS}
    
    Electron and positron ($e^{\pm}$) fluxes are generated in DM annihilations
mainly through the hadronization, fragmentation and decays of the annihilation products, since direct production of $e^+ e^-$ is supressed by small couplings in most DM models. These $e^{\pm}$ propagate in the Galaxy and produce high-energy gamma rays via ICS off the ambient photon background. 
The differential flux produced by ICS from a given observational region $\Delta\Omega$ in the Galactic halo is given by \cite{Cirelli:2010xx}  
\begin{eqnarray}
\frac{d\Phi^{ICS}_\gamma}{dE_\gamma}&=&\sum_i\frac{\langle\sigma_i v \rangle}{8\pi m_{DM}^2}\int_{m_e}^{m_{DM}}\frac{dE_I}{E_\gamma}
\frac{dN^{i}_{e^\pm}}{dE_e}(E_I)\int d\Omega\ \frac{I_{IC}(E_{\gamma},E_I;\Psi)}{E_\gamma}\ ,
\label{gammafluxICS}
\end{eqnarray}
where $E_I$ is the $e^{\pm}$ injection energy, $\Psi$ corresponds to the angular position where the ICS gamma rays are produced, and the function $I_{IC}(E_{\gamma},E_I;\Psi)$ is given by
\begin{eqnarray}
I_{IC}(E_{\gamma},E_I;\Psi)&=&2E_{\gamma}\int_{l.o.s.}dl
\int_{m_e}^{E_I}dE_e\ \frac{P_{IC}(E_\gamma,E_e;{\bf x})}{b_{T}(E_e;{\bf x})}\ \tilde I(E_e,E_I;{\bf x})\ .
\label{gammafluxICS2}
\end{eqnarray}
Here ${\bf x}=(l,\Psi)$ and $b_{T}\propto E^2$ is the energy-loss rate of the electron in the Thomson limit. The function $P_{IC}$ is the photon emission power for ICS, and it depends on the interstellar radiation (ISR) densities for each of the species composing the photon background. It is known that the ISR in the inner Galactic region can be well modeled as a sum of separate black body radiation components corresponding to star-light (SL), infrared radiation (IR), and cosmic microwave background (CMB) \cite{Cirelli:2009vg}.  In this work we have used  the interstellar radiation field provided by {\tt GALPROP} \cite{galprop} to calculate the normalization and the temperature for each of these three components.  For the injection spectra of $e^\pm$, we utilize pre-evaluated tables in \cite{Cirelli:2010xx}.

The last ingredient in Eq. (\ref{gammafluxICS2}) is the $\tilde I(E_e,E_I;{\bf x})$ function, which can be given in terms of the well- known halo function \cite{Cirelli:2010xx}, $I(E ,E_I;{\bf x})=\tilde I(E ,E_I;{\bf x})[(b_T(E)/b(E,{\bf x}))(\rho({\bf x})/\rho_\odot)^2]^{-1}$, where $\rho_\odot$ is the DM density at Sun's position and $b(E,{\bf x})$ encodes the energy loss of the $e^\pm$. The $\tilde I(E_e,E_I;{\bf x})$ function obeys the diffusion loss equation \cite{Cirelli:2010xx}, 

\begin{equation}
\nabla^2 \tilde I(E_e,E_I;{\bf x}) + \frac{E_e^2}{K(E_e;{\bf x})}\frac{\partial}{\partial E_e}\left[\frac{b(E_e;{\bf x})}{E_e^2} \tilde I(E_e,E_I;{\bf x})\right]=0\ ,
\label{difflosseq}
\end{equation}
and is commonly solved by modeling the diffusion region as a cylinder with radius $R_{\rm max} = $20 kpc, height $z$ equal to $2L$ and vanishing boundary conditions.  Also the diffusion coefficient $K(E;{\bf x})$ has been taken as homogeneous inside the cylinder with an energy dependence following a power law $K(E)=K_0 (E/1{\rm GeV})^\delta$.
For these three parameters $L$, $K_0$ and $\delta$, the so called diffusion coefficient, we have adopted three sets referred to as MIN, MED and MAX models \cite{Delahaye:2007fr}, which account for the degeneracy given by the local observations of the cosmic rays at the Earth including the boron to carbon ratio, B/C \cite{Maurin:2001sj}. We take them as our benchmark points, although we note that MIN and MAX models do not imply minimal or maximal expected gamma-ray signal, respectively. To solve this equation under the described conditions, we have used {\tt BoxLib} \cite{boxlib} which is a general purpose partial differential equation solver with an adaptive mesh refinement method.
We show in Section \ref{limits} that the use of the different diffusion models, MIN, MED, MAX, does not introduce a large variation in the
DM constraints.

Let us finally remark about the importance of the energy loss function $b(E;{\bf x})$. The two main energy loss mechanisms of $e^\pm$ in the Galaxy are the ICS and synchrotron radiation  produced by interaction with the Galactic magnetic field. The former is the only contribution to the energy losses that is usually considered, since it is the most important one in studies of sources far from the GC. But when the $e^\pm$ energy reaches several hundreds of GeV, synchrotron radiation can dominate the energy loss rate due to the suppression factor in the ICS contribution in the Klein-Nishina regime. By contrast, synchrotron radiation losses do not have this suppression, and are driven by the magnetic field energy density  $u_B({\bf x})=B^2/2$.  Although the strength and exact shape of the Galactic magnetic field is not well known, in the literature it is broadly described by the from \cite{Cirelli:2010xx}, 
\begin{equation}
B(r,z)=B_0\ {\rm exp}\left(-\frac{r-8.5\ {\rm kpc}}{10\ {\rm kpc}}-\frac {z}{2\ {\rm kpc}}\right)\ , 
\label{magnetico}
\end{equation}
normalized with the strength of the magnetic field around the solar system, $B_0$, which is known to be in the range of 1 to $10\ \mu{\rm G}$ \cite{galprop}. This field grows towards the GC and therefore one should expect that the energy losses are dominated by synchrotron radiation in the inner part of the galaxy \cite{Cirelli:2010xx}. 
On the other hand, we can expect that when the magnetic field is stronger, the energy of the injected $e^\pm$ is more efficiently liberated in
the form of synchrotron emission resulting in a softer spectrum, and producing therefore smaller constraints on the DM annihilation cross-section.
We will check this in Section \ref{limits}. There we will also check that the ICS effect becomes relevant only for the
$\mu^+ \mu^-$ annihilation channel, since the contribution of the prompt gamma rays is less important than in other channels with hadronic decays.

    \begin{figure}[!t]
      \centering
	\includegraphics[width=0.28\textwidth]{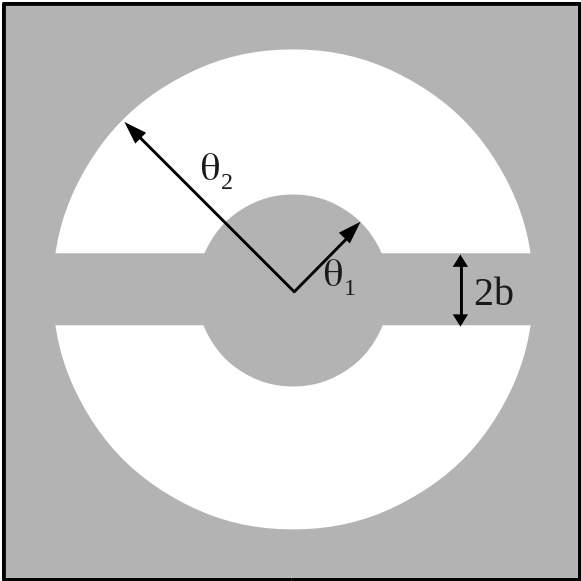}

	 \caption{\label{fig:data_roi} Schematic view of our choice of the ROI. The gray area represents the masked region. 
} 
      \end{figure}

\section{Gamma-ray flux from \textit{Fermi}-LAT measurements 
}

\label{data}

  \subsection{Data selection and processing}
	   
	      The Fermi satellite was launched on June $11$, 2008. Its main instrument, the Large Area Telescope (LAT) \cite{Atwood:2009ez}, 
	      collects high energy gamma rays ($\sim 20$ MeV to $>300$ GeV) with a large effective area ($\sim 6200$ cm$^2$ above 
	      $1$ GeV for P7CLEAN$\_$V6 at normal incidence \cite{Ackermann:2012uf}) and a large field of view ($2.4$ sr).
	      Further details on the LAT can be found in Ref.~\cite{Atwood:2009ez,Ackermann:2012uf}.

	      In our analysis we use the LAT photon data measured between August 4, 2008, and June 15, 2012, in the energy range between $1$ GeV and $100$ GeV. 
	      Events with zenith angles $<100^{\circ}$ are selected to reduce the contamination by gamma-ray emission coming 
	      from cosmic-ray interactions in the atmosphere.
We select events from the P7ULTRACLEAN\_V6::FRONT class. This choice reduces the cosmic-ray background contamination and takes advantage of a narrower PSF
	      as compared to back-converting events. We make a reasonable assumption on systematic uncertainty extending it from Source and Clean classes. The systematic uncertainty of the effective area for both Source and Clean class
	      events is quoted as $10\%$ at $100$ MeV, decreasing to $5\%$ at $560$ MeV, and increasing to $20\%$ at $10$ GeV and above \cite{Ackermann:2012uf}. Maps of flux for different energy ranges from a region of $30^{\circ}$ around the GC is made using version V$9$r$28$ of the LAT Science Tools \cite{science_tools}. As we will show below, we can use a single flux map (built summing up the flux maps for the different energy ranges) 
	      to select the Region Of Interest (ROI) with the aim of maximizing the $S/N$ ratio for each individual DM profile under study.

\subsection{ Optimization of the region of interest for dark matter searches} \label{ROI}
\begin{figure}[!b]
      \centering
        \hspace*{-1.5cm}
        \includegraphics[width=0.4\textwidth]{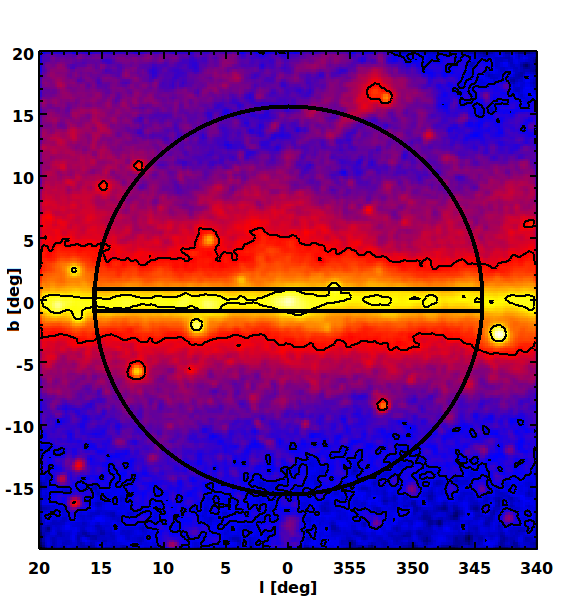}
        \includegraphics[width=0.4\textwidth]{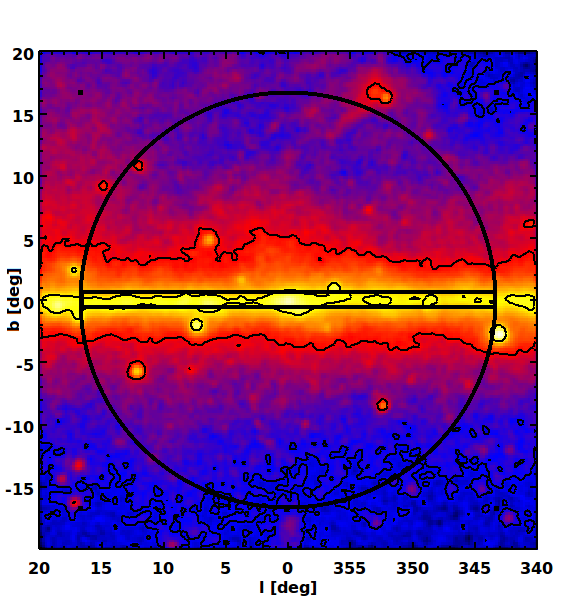}
        \\
        \vspace*{0.5cm}
        \hspace*{-1.5cm}
        \includegraphics[width=0.4\textwidth]{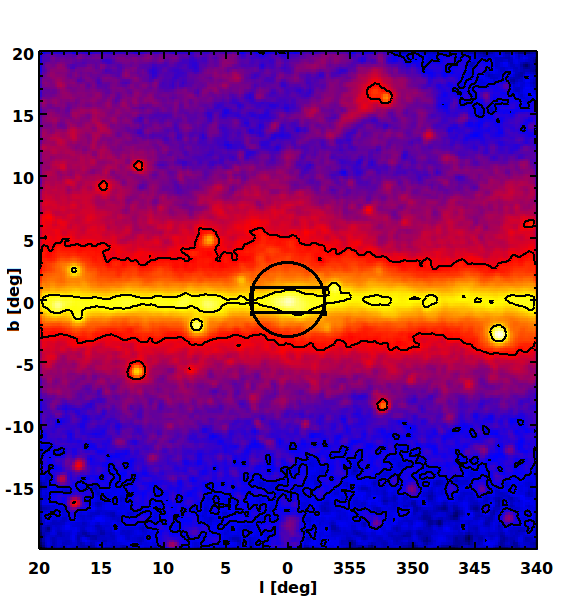}
        \includegraphics[width=0.4\textwidth]{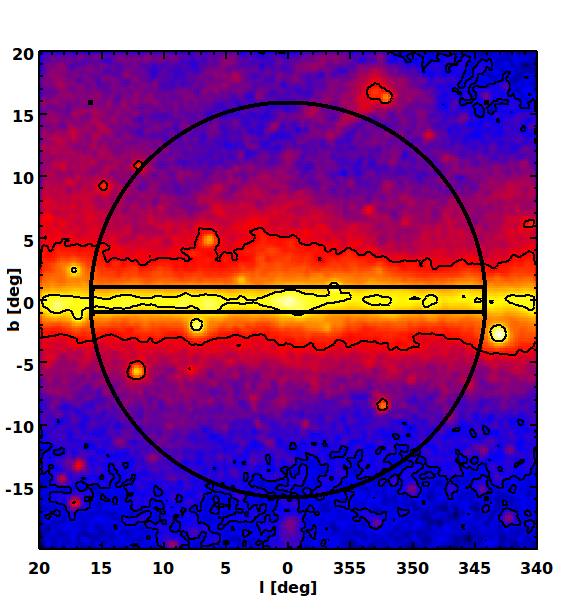}     
          \caption{\label{fig:data_roi_color} 
Maps of the observed flux by the \textit{Fermi}-LAT in the energy range $1-100$ GeV, in units of photons cm$^{-2}$ s$^{-1}$, for the four DM profiles studied. Upper left: Einasto, upper right: NFW, bottom left: NFW$_c$, and bottom right: Burkert.
For each profile, the ROI is the region inside the circle excluding the band on the Galactic plane.
Color scale is logarithmic, yellow, red and blue correspond to $3.6\times10^{-9}$, $6.4\times10^{-10}$ and $1.2\times10^{-10}$ 
photons cm$^{-2}$ s$^{-1}$, respectively. These values also correspond to black contours. In order to reduce statistical noise and to bring up finer features in the inner galaxy the map is smoothed with a $0.2^{\circ}$ FWHM Gaussian function.}
   \end{figure}

             An important step in our analysis is the optimization of the ROI using a data-driven procedure 
             that maximizes the $S/N$ ratio. In order to do so, we follow a procedure similar to the one presented in the appendix 
             A of \cite{Bringmann:2012vr}:

             \begin{enumerate}
              \item  We produce $40^{\circ}\times 40 ^{\circ}$ maps centered on the GC of the quantity $\bar J (\Delta\Omega)\Delta\Omega$  for the four 
              DM density profiles considered (i.e., Einasto, NFW, NFWc and Burkert) and use them as signal. 
              Each pixel $i$ has an area of $0.2^{\circ}\times 0.2^{\circ}$  and contains a J-factor value $J_i$ calculated with Eq.~(\ref{gammaflux2}).
              \item We use as noise the square root of the observed photon flux integrated in the energy range 1-100 GeV. We use a single map, instead of a different one for each energy bin since the morphology of the background does not exhibit strong variations in energy. The flux in pixel $i$ is labelled as $F_i$.
              \item A mask, defined by three angles $\theta_1$, $\theta_2$ and $|b|$ as shown in Figure \ref{fig:data_roi}, is introduced 
                    to cover the GC, the Galactic plane, and the high galactic latitude regions, where the statistical fluctuations of the Galactic foreground dominate over the expected DM signal.
              \item The optimization procedure consists of finding the set of angles that mask a region such that the $S/N$ 
	            ratio is the largest for each DM profile considered. What we technically do is to minimize the inverse of the following quantity
	            \begin{eqnarray}\label{quanty}
	              S/N = \frac{\sum_i{J_i}}{\sqrt{\sum_i{F_i}}}
	            \end{eqnarray}
	            with $i$ running over unmasked pixels, varying masks. We use the numerical routine {\tt Fmin} of the Python module {\tt scipy.optimize} \footnote{http://docs.scipy.org/doc/scipy/reference/optimize.html\#}, which minimizes a function using the downhill simplex algorithm. 
	            We end up with four masks characterized by those angles given in Table \ref{tab:table2}.
In the same Table, we also show the values of $\Delta\Omega$ and $\bar J (\Delta\Omega)\Delta\Omega$ for each profile. 
	            As expected, $\theta_1 = |b|$ for all the profiles, since the broadest emission in the Galactic plane is the one in the GC.
	            
             \end{enumerate}       
Figure \ref{fig:data_roi_color} shows the ROI that we have obtained for each DM density profile in Table \ref{tab:table1}. Clearly, the NFW$_c$ ROI is the smallest one. 
This can be easily understood by inspecting the right panel of Figure~\ref{fig:J}: the $\bar J (\Delta\Omega)\Delta\Omega$ quantity for NFW$_c$ becomes almost constant beyond only $5^{\circ}$, whereas for the other profiles this quantity becomes flat at much larger radii. Therefore, in the case of the NFW$_c$ profile, increasing the angular aperture by a few more degrees does not increase the $S/N$.

Note that the usual quantities to calculate $S/N$ ratios are observed counts and expected DM-induced counts but in this work, instead, we use observed gamma-ray flux (rather than counts) and J-factors (formally proportional to the expected DM-induced gamma-ray flux). 
Nevertheless, we checked that the use of observed fluxes and predicted J-factors turns out to be a very good approximation, which leads to similar optimized ROIs. We performed the following test. Using the gtobssim tool and assuming the NFW profile, we simulated the events that different DM models could produce in the LAT after 46 months of observation. We used the same IRFs, cuts and procedure to select events as those used for the real observations. Instead of using a fixed $1-100$ GeV energy range we further optimize this quantity choosing an energy range centered around the DM emission peak. We then compare the simulated DM counts maps and the actually observed counts map in the given energy range to re-derive the optimized $S/N$ region. As anticipated above we find that the derived ROI’s parameters using counts maps do not change significantly for different DM models from those calculated using observed fluxes and J-factors and a fixed energy range.
\begin{table}[!t]
\centering
\begin{tabular}{|c|c|c|c|c|c|c|}
   \hline
  Profile	\rule{0pt}{3ex}   & $\theta_1$ & $\theta_2$ & $|b|$& $\Delta\Omega$ &         $\bar J\left(\Delta\Omega\right)\Delta\Omega$           &  Flux ($1-100$ GeV)    \\
  &  [$^{\circ}$]     &   [$^{\circ}$]    & [$^{\circ}$]&      [sr]      &    [$\times 10^{22}$ GeV$^2$ cm$^{-5}$ sr]         & [$\times 10^{-7}$ cm$^{-2}$s$^{-1}$] \\
   \hline \rule{0pt}{3ex} 

  Burkert \rule{0pt}{3ex} &    $0.8$   &    $15.9$   & $0.8$&     0.225      &          $41.9$                            &           $32.1 \pm 0.3$      \\
  Einasto  &       $0.7$   &    $15.6$   & $0.7$&     0.217      &          $5.1$                            &           $31.4 \pm 0.3$ \\ 
   NFW   &    $0.6$   &    $16.7$   & $0.6$&     0.253      &            $3.3$                            &          $38.0 \pm 0.3$        \\
   NFW$_c$  &   $1.0$   &    $ 3.0$   & $1.0$&     0.005      &             $86.8$                         &     $ 2.2 \pm 0.1$             \\ 
   \hline
\end{tabular}
\caption{\label{tab:table2} The optimized regions for the DM density profiles studied, defined by the angles shown in Figure \ref{fig:data_roi}.
The corresponding values for $\Delta\Omega$, $\bar J (\Delta\Omega)\Delta\Omega$, and observed flux with statistical errors only
 (in the energy range between $1-100$ GeV) are also shown. 
}
\end{table}

  \subsection{Flux measurement}
  
\begin{figure}[!t]
      \centering
	\includegraphics[width=0.7\textwidth]{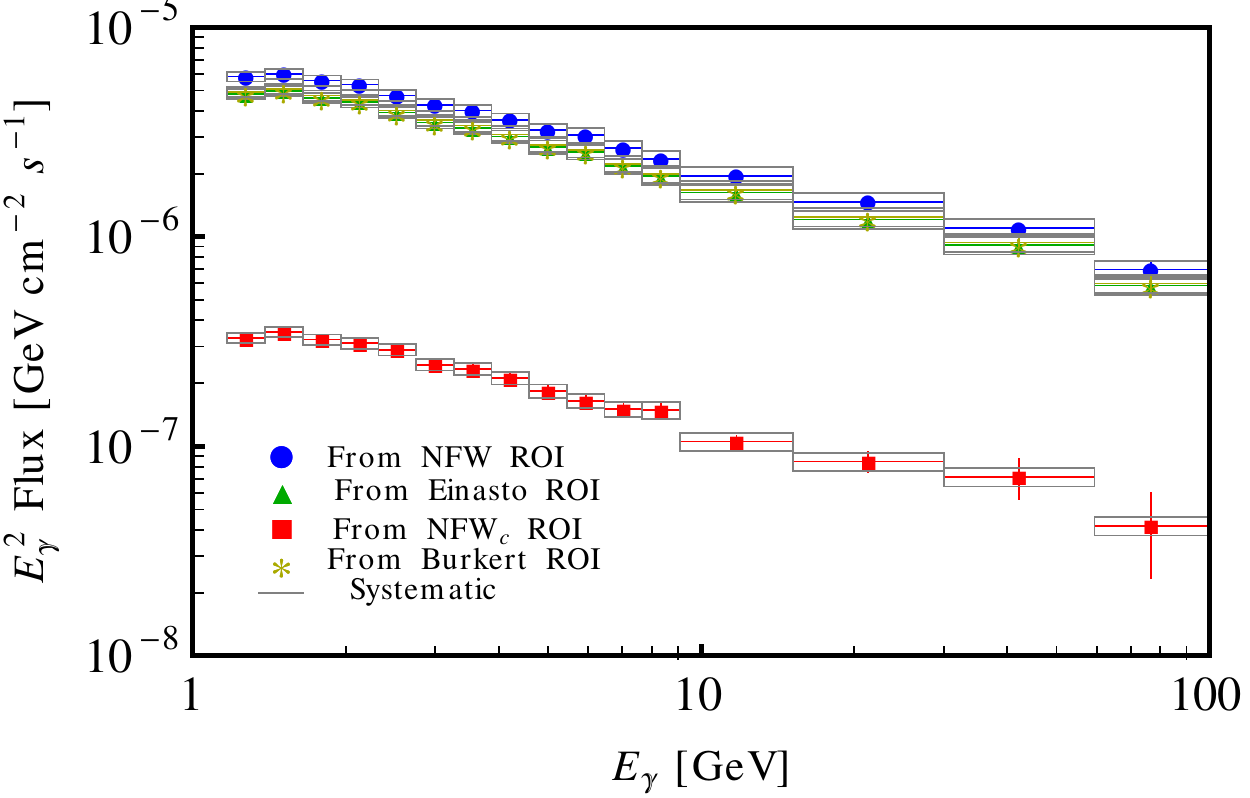}
	 \caption{\label{fig:spectrum} Energy spectra extracted from \textit{Fermi}-LAT data for the optimized regions that are shown in Figure~\ref{fig:data_roi_color}.
Data are shown as points and the vertical error bars represent the statistical errors. The latter are in many cases smaller than the point size.
The boxes represent the systematic error in the \textit{Fermi}-LAT effective area. 
} 
	  \end{figure}

Following the analysis described above, we show in Figure \ref{fig:data_roi_color} the flux observed by the \textit{Fermi}-LAT, and the ROIs corresponding to each of the DM profiles considered. The value of this flux integrated in the energy range $1-100$ GeV can be found in the last column of Table \ref{tab:table2}. The energy spectra from the ROI associated to each profile are shown in Figure \ref{fig:spectrum}. We limit the energy range of the analysis to be below $100$ GeV in order to have a small statistical uncertainty in each bin, falling generally below the systematic uncertainty. In this way we remove the possibility for the upper limits to be accidentally dominated by a large downward fluctuation in the energy bins close to the peak of the gamma-ray emission from DM annhilation, which is the most constraining point when comparing to the measured flux.

 
To set constraints we require that the DM-induced gamma-ray flux does not exceed the flux upper limit (UL) evaluated as follows. We set $99.98\%$ UL signal counts using the Bayesian approach presented in Ref.~\cite{Mazziotta}, for the case of absence of background with systematic uncertainties not included, which correctly takes into account the Poisson limit (i.e. the case of low counts).
Using exposure maps calculated with the gtexpcube2 tool of the Science Tools we are able to convert UL signal counts into the UL signal flux needed to set constraints.

\section{Limits on the dark matter annihilation cross-section}

\label{limits}


As already discussed, we adopt a conservative approach in the analysis of the limits on the DM annihilation cross section, simply 
requiring that the integrated gamma-ray flux of the expected DM signal for each energy bin does not exceed the 
upper limit signal flux evaluated following the Bayesian procedure in Ref.~\cite{Mazziotta}\footnote{It is worth noting that even though we optimize the ROIs based on both, DM and observed distributions, to set limits on DM annihilation cross section we perform a spectral analysis. It would be interesting for a future work to check that at the upper limit cross sections derived here, the implied spatial distribution of the gamma-ray signal intensity does not significantly exceed the data anywhere within the ROI at any energy.}.
We do not subtract any astrophysical background.  

\begin{figure}[!t]
\centering
\hspace*{-1.5cm}
	\includegraphics[width=0.4\textwidth]{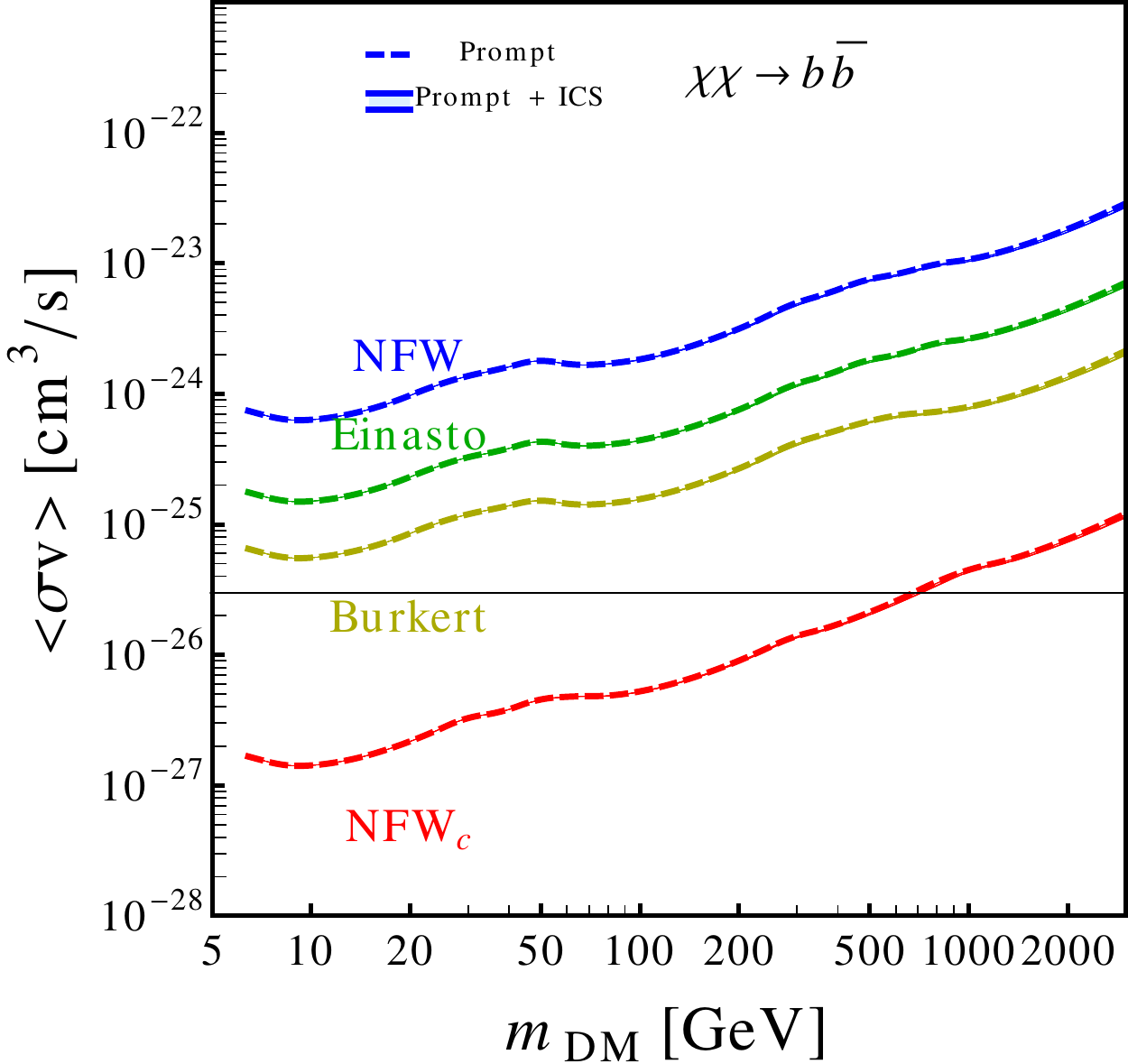}
        \includegraphics[width=0.4\textwidth]{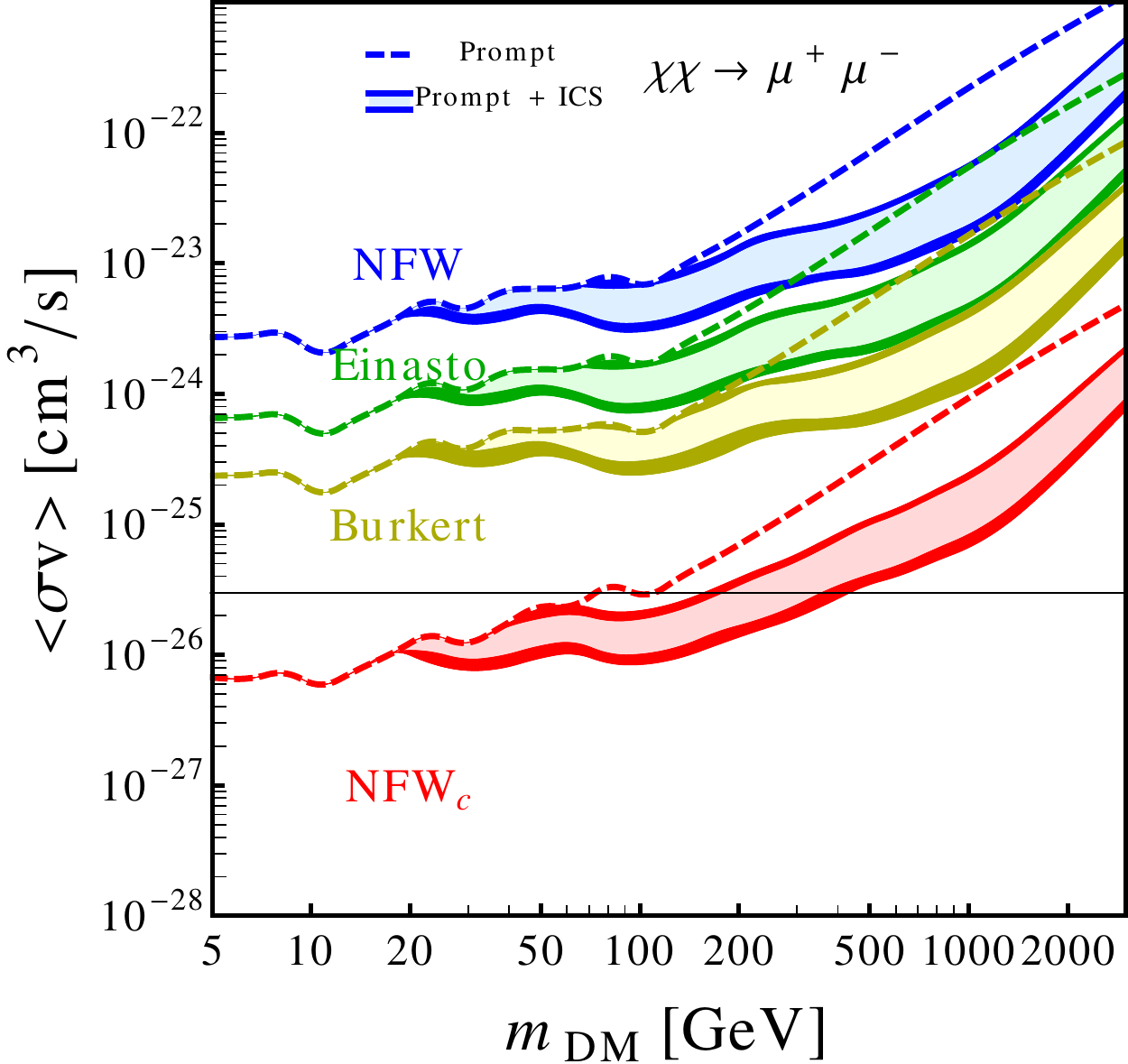}
        \\
        \vspace*{0.5cm}
        \hspace*{-1.5cm}
 	\includegraphics[width=0.4\textwidth]{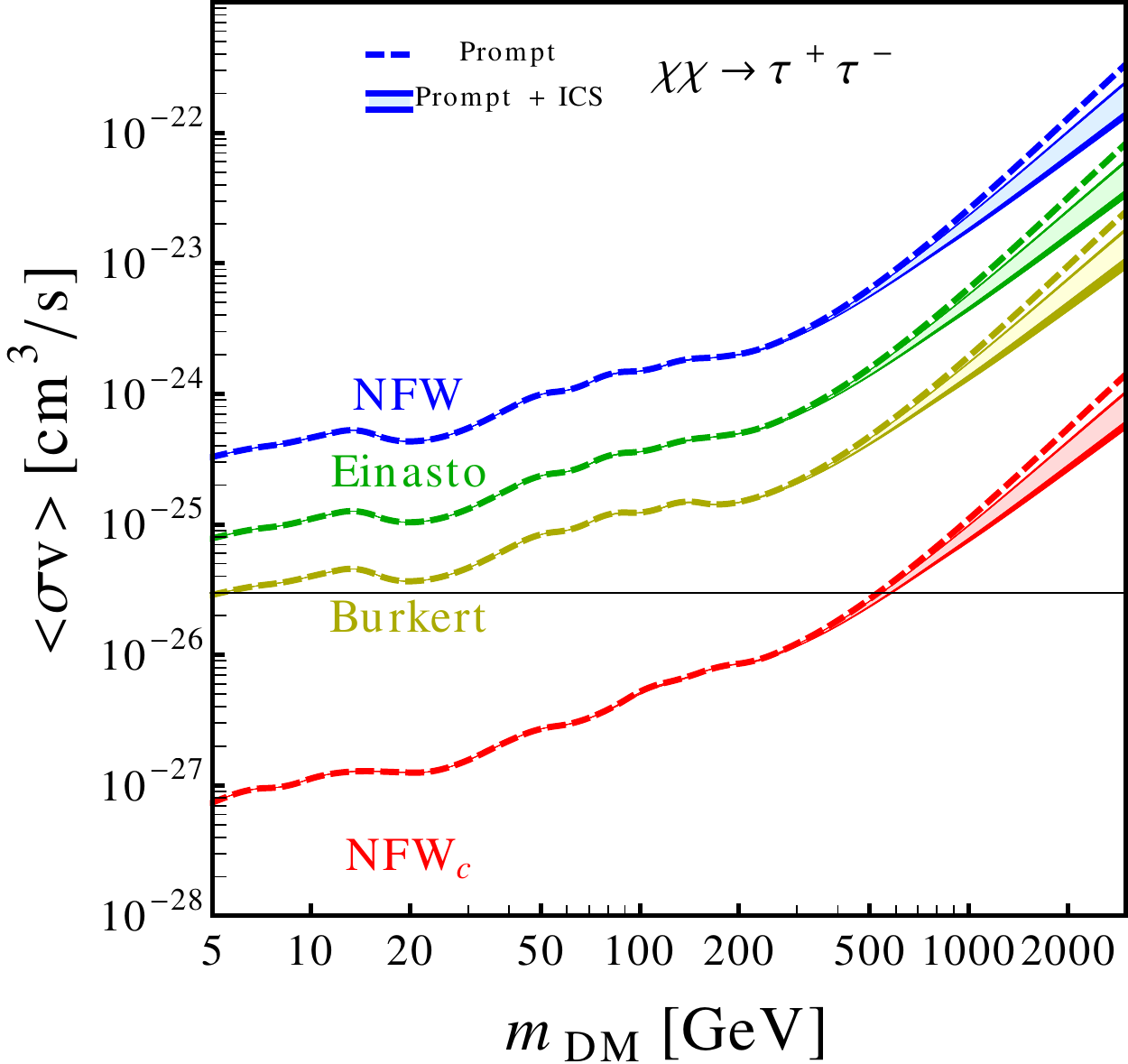}
 	\includegraphics[width=0.4\textwidth]{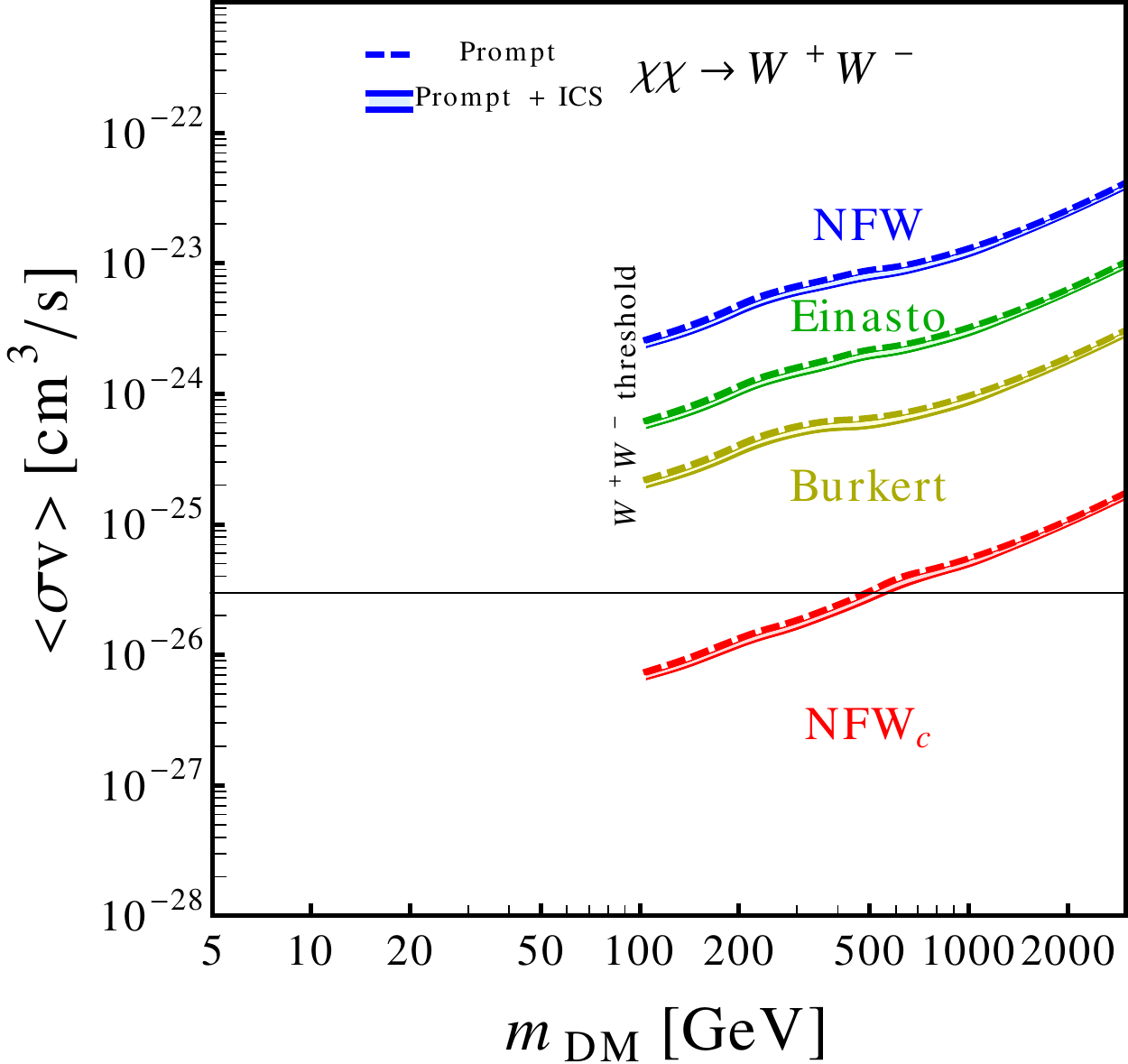}
	 \caption{\label{fig:const_b} $3\sigma$ upper limits on the annihilation cross-section of models in which DM annihilates into $b\bar b$, 
$\mu^+\mu^-$ (upper panel), $\tau^+\tau^-$ or $W^+W^-$ (lower panel), for the four DM density profiles discussed in the text.
Upper limits set without including the ICS component in the computation are also given as dashed curves (prompt) for comparison.
The uncertainty in the diffusion model is shown as the thickness of the solid curves (from top to bottom: MIN, MED, MAX) while the lighter shaded regions represent the impact of the different strengths of the Galactic magnetic field with lower(higher) values of the cross-section corresponding 
to $B_0=1~\mu$G($B_0=10~\mu$G). 
The horizontal line corresponds to the expected value of the thermal cross-section for a generic WIMP candidate.} 
        \end{figure}

We present the results in Figure~\ref{fig:const_b}, where the constraints obtained are shown for different final states.
There we also illustrate
the case 
$\langle\sigma v\rangle=3\times 10^{-26}$ cm$^3$ s$^{-1}$, which corresponds to the value of the annihilation cross-section 
associated to the correct thermal relic abundance 
for a WIMP whose annihilation is dominated by the s-wave (velocity-independent) contribution and thus,
$\Omega_{DM}~h^2\approx 
3\times 10^{-27}$ cm$^3$ s$^{-1}\ \langle\sigma v\rangle ^{-1}\approx 0.1$~\cite{jungman96}. For comparison, the constraints are given considering only the contribution from prompt gamma rays and the total contribution from 
prompt plus ICS gamma rays.

First, it is worth noting that if the DM density follows an Einasto, NFW or Burkert profile, the upper limits on the annihilation cross
section are above the value of the thermal cross-section for any annihilation channel.
Nevertheless, the situation is drastically different
when we consider the DM compression due to baryonic infall in the inner region of the Galaxy. Indeed, by adopting the NFW$_c$ profile 
and for a
$b\bar b$, $\tau^+\tau^-$ and $W^+W^-$ channel,
the thermal annihilation cross-section is already reached for a DM mass of 
680, 530 and 490 GeV, respectively.
For the $\mu^+ \mu^-$ channel the effect of the prompt gamma rays is less important 
since generally fewer
photons are produced in the FSR compared to the hadronic decays of the other channels.
(For the $W^+W^-$ which is open when $m_{DM} \gtrsim 90$ GeV, the 
 $W^{\pm}$ decays produce a large number of photons, especially at high energy).
Notice that the lower bound associated with prompt gamma rays for $\mu^+ \mu^-$ is 100 GeV compared to about 500--700 GeV in the other channels.
Thus the ICS is 
important in this case, also due to the relatively harder $e^\pm$ spectrum \cite{Birkedal:2005ep}.
We can see that
for $B_0=1~\mu$G the lower bound on the DM mass turns out to be 358 GeV and for $B_0=10~\mu$G the bound is 157 GeV, using the 
MIN diffusion model. For MED and MAX diffusion models the values turn out to be 404, 171 GeV and 439, 179 GeV, respectively.
As discussed in Subsection \ref{ICS}, when the magnetic field is stronger the energy of the injected $e^\pm$ is more efficiently liberated in the form of microwaves, resulting in a softer gamma-ray spectrum, and producing therefore lower constraints . Therefore, we have shown that in those cases in which the ICS component is dominant (for heavy WIMP masses in general), the variation of the magnetic field can significantly alter the expected gamma-ray fluxes from the inner regions of the Galaxy.

Although the above results can be interpreted in general as implying that vanilla WIMP models and contracted DM profiles are incompatible with the Fermi data, one should keep in mind that if one works in the framework of a specific particle physics model this conclusion might in principle be avoided in some regions of the parameter space. For example, the final state can be a combination of the annihilation channels presented here, as in supersymmetry where the lightest neutralino annihilation modes are $70\% ~\overline b b~-~ 30\% ~\overline \tau \tau$ for a Bino DM, and $100\% ~W^+ W^-$  for a Wino DM (or for a Higgs-portal model). More importantly, the value of the annihilation cross section in the Galactic halo might be smaller than $3\times 10^{-26}$ cm$^3$ s$^{-1}$ for a DM candidate that is thermally produced. For example, in the early Universe coannihilation channels can also contribute to $\langle\sigma v\rangle$.
Also, DM particles whose annihilation in the early Universe is dominated by p-wave (velocity-dependent) contributions would have a smaller value of $\langle\sigma v\rangle$ in the Galactic halo, where the DM velocity is much smaller than at the time of freeze-out, and can therefore escape the constraints derived in this work.
These two effects can in fact occur in some regions of the parameter space of well motivated models for particle DM, such as the neutralino. In this sense, the results derived above for pure annihilation channels can be interpreted as limiting cases that give an idea of what can happen in realistic scenarios.

Let us remark that the upper limits on the annihilation cross-section that we have obtained for the cases of NFW, Einasto and Burkert
profiles are comparable to the ones previously reported 
by the \textit{Fermi}-LAT collaboration \cite{fermidiffuse}, after a similar analysis of the Galactic halo without modeling of the astrophysical background
(similar results were also obtained in Ref.~\cite{fermidiffuse2, fermidiffuse3}). 
Modeling of the background was also considered in
Ref.~\cite{fermidiffuse}, and the results
are competitive with those from dSphs \cite{Ackermann:2011wa, Abdo:2010, GeringerSameth:2011iw}, where the upper limit of the
annihilation cross section is below the thermal one for DM masses
smaller than 27 and 37 GeV assuming a $b\bar b$ and a $\tau^+\tau^-$ channel, respectively.  
Remarkably, when we take into account the baryonic infall in our conservative analysis, forcing the DM to contract in those inner regions of the Galaxy, we obtain much stronger limits. In particular, as discussed above, using our compressed DM density profile, NFW$_c$, the thermal cross section is excluded for a DM mass smaller than 680 and 530 GeV in the $b\bar b$ and $\tau^+\tau^-$ channel, respectively,
thus improving those limits obtained from dSphs \cite{Ackermann:2011wa, Abdo:2010, GeringerSameth:2011iw}, and also those obtained from galaxy clusters \cite{frenk}. In the latter, DM masses smaller than about 100 GeV are constrained provided that DM subhalos significantly contribute to boost 
the DM signal.
 
In general,
the upper limits on the DM annihilation cross section are two orders of magnitude stronger than without contraction.
In recent works on the GC a similar analysis was carried out but subtracting the emission from known point sources 
and from the Galactic disk \cite{hooper, kelso}.


\section{Conclusions}

\label{conclusions}

We derived constraints on the parameter space of generic DM candidates
using \textit{Fermi}-LAT inner Galaxy measurements.  We
considered well motivated DM density profiles, such as Burkert,
Einasto and NFW, which are perfectly compatible with current observational data of the Milky Way.
We then selected optimal regions around the GC, such that the $S/N$ ratio is
maximized. When the effect of contraction in the DM halo due to
baryons is included in the computation \cite{Prada:2004pi,mambrini},
the constraints turn out to be very strong. 
In particular, a compressed DM density profile allows us to place upper limits on the DM annihilation cross section that exclude
the thermal cross section for a broad range of DM masses,
as shown in Figure~\ref{fig:const_b}. This is the
case for masses smaller than 680, 530 and 490 GeV for $b\bar b$, $\tau^+\tau^-$
and $W^+W^-$ channels, respectively.  For the $\mu^+ \mu^-$ channel, where the ICS
effect is important, the exclusion of the thermal
cross-section is for a mass smaller than about 150 to 400 GeV, depending on models of the Galactic magnetic field. Alternatively, one may interpret these results as implying that vanilla WIMP models
 and contracted DM profiles are incompatible with the \textit{Fermi} data.

Although the constraints are very strong,
the analysis is conservative since we require that the
expected DM signal does not exceed the gamma-ray emission observed by the \textit{Fermi}-LAT, and modeling of the astrophysical background is not
carried out. The latter would only lead to better constraints on the DM annihilation cross section.








\section*{Acknowledgments}

We gratefully acknowledge A. Cuoco, G. Zaharijas and L. Latronico for valuable comments and helpful discussions during the preparation of the manuscript.

This work was supported by the Spanish MINECO's Consolider-Ingenio 2010 
Programme under grant MultiDark CSD2009-00064. 
The work of DGC, GAGV, JHH, CM and MP was supported in part by MINECO under grants
FPA2009-08958 and FPA2012-34694, 
and under the `Centro de Excelencia Severo Ochoa' Programme SEV-2012-0249, 
by the Comunidad de Madrid under grant HEPHACOS S2009/ESP-1473, 
and by the European Union under the Marie Curie-ITN program PITN-GA-2009-237920. 
JHH was also supported by DOE grant DE-FG02-13ER42022.
The work of YM was supported by  the French ANR TAPDMS ANR-09-JCJC-0146
and acknowledge partial support from the European Union FP7 ITN INVISIBLES (Marie
Curie Actions, PITN- GA-2011- 289442).
GAGV thanks Caltech and SLAC for hospitality
during the completion of this work.

The \textit{Fermi} LAT Collaboration acknowledges generous ongoing support
from a number of agencies and institutes that have supported both the
development and the operation of the LAT as well as scientific data analysis.
These include the National Aeronautics and Space Administration and the
Department of Energy in the United States, the Commissariat \`a l'Energie Atomique
and the Centre National de la Recherche Scientifique / Institut National de Physique
Nucl\'eaire et de Physique des Particules in France, the Agenzia Spaziale Italiana
and the Istituto Nazionale di Fisica Nucleare in Italy, the Ministry of Education,
Culture, Sports, Science and Technology (MEXT), High Energy Accelerator Research
Organization (KEK) and Japan Aerospace Exploration Agency (JAXA) in Japan, and
the K.~A.~Wallenberg Foundation, the Swedish Research Council and the
Swedish National Space Board in Sweden.

Additional support for science analysis during the operations phase is gratefully
acknowledged from the Istituto Nazionale di Astrofisica in Italy and the Centre National d'\'Etudes Spatiales in France.

\end{document}